\documentclass[colorlinks=true,linkcolor=black,urlcolor=black,citecolor=blue,submission,copyright,creativecommons]{eptcs}

\providecommand{\event}{DCM 2011}

\usepackage{amsmath,amsfonts,amssymb,amsthm}
\usepackage{prooftree}
\usepackage{graphicx}

\newtheorem{theorem}{Theorem}[section]
\newtheorem{lemma}[theorem]{Lemma}
\newtheorem{corollary}[theorem]{Corollary}

\newtheorem{example}{Example}[section]

\newcommand{\lvec}{\ensuremath{\lambda^{\!\!\textrm{vec}}}}
\newcommand{\canon}[1]{[#1]}
\newcommand{\cocanon}[1]{\{#1\}}

\newcommand{\Bool}{\mathbb{B}}
\newcommand{\True}{\mathbb{T}}
\newcommand{\False}{\mathbb{F}}
\newcommand{\true}{{\bf true}}
\newcommand{\false}{{\bf false}}
\newcommand{\eg}{\emph{e.g.}~}
\newcommand{\ie}{\emph{i.e.}~}

\newcommand{\ve}[1]{\mathrm{\textbf{#1}}}
\newcommand{\type}{\colon\!}
\newcommand{\Sc}{\mathsf{S}}
\newcommand{\sui}[1]{\sum_{i=1}^{#1}}
\newcommand{\suj}[1]{\sum_{j=1}^{#1}}

\newcommand{\ket}[1]{{|{#1}\rangle}}

\newcommand{\Neutral}{\mathcal{N}}
\newcommand{\CT}{\Lambda_0}
\newcommand{\SN}{{\it SN}_0}
\newcommand{\Red}{{\rm Red}}
\newcommand{\RC}{\mathsf{RC}}
\newcommand{\RCn}{{\bf RC}}
\newcommand{\bcal}[1]{\mathsf{#1}}
\newcommand{\denot}[1]{{[\!|{#1}|\!]}}

\title{A Type System for the Vectorial Aspect of the Linear-Algebraic Lambda-Calculus}
\author{Pablo Arrighi
\institute{LIP, \'{E}cole Normale Sup\'erieure de Lyon\\
46 all\'ee d'Italie\\
69364 Lyon cedex 07, France
}
\institute{LIG, Universit\'e de Grenoble\\
220, rue de la Chimie\\
38400 Saint Martin d'H\`eres, France}
\email{pablo.arrighi@imag.fr}
\and
Alejandro D\'iaz-Caro
\institute{LIG, Universit\'e de Grenoble\\
220, rue de la Chimie\\
38400 Saint Martin d'H\`eres, France}
\institute{LIPN, Universit\'e Paris 13, Sorbonne Paris Cit\'e\\
99 av. J-B Cl\'ement,\\
93430 Villetaneuse, France}
\email{alejandro@diaz-caro.info}
\and
Beno\^it Valiron
\institute{LIPN, Universit\'e Paris 13, Sorbonne Paris Cit\'e\\
99 av. J-B Cl\'ement,\\
93430 Villetaneuse, France}
\institute{University of Pennsylvania\\
CIS Department\\
Philadelphia, PA 19104, USA}
\email{benoit.valiron@monoidal.net}
}

\def\titlerunning{A Type System for the Vectorial Aspects of the
  Linear-Algebraic
  $\lambda$-Calculus}
\def\authorrunning{P. Arrighi, A. D\'iaz-Caro \& B. Valiron}
\begin{document}
\maketitle

\begin{abstract}
  We describe a type system for the linear-algebraic
  lambda-calculus. The type system accounts for the part of the
  language emulating linear operators and vectors, i.e. it is
  able to statically describe the linear combinations of terms
  resulting from the reduction of programs.  This gives rise to an
  original type theory where types, in the same way as terms, can be
  superposed into linear combinations.  We show that the resulting
  typed lambda-calculus is strongly normalising and features a weak
  subject-reduction.
\end{abstract}

\section{Introduction}
A number of recent works seek to endow the $\lambda$-calculus with a
structure of vector space; this agenda has emerged simultaneously in
two different contexts (albeit
related~\cite{DiazcaroPerdrixTassonValironHOR10}). A first line of
work forked from the study of relational models of linear logic. In
\cite{EhrhardRegnierTCS03,TassonTLCA09,VauxMSCS09}, various algebraic
lambda-calculi, that is, languages with vectorial structures, are
considered. These languages are based on an interpretation of
intuitionistic logic by linear logic. A second line of work
\cite{ArrighiDiazcaroQPL09,ArrighiDowekRTA08,DiazcaroPetitWoLLIC12}
considers linear combinations of terms as a sort of ``quantum
superposition''. This paper stems from this second approach.

In quantum computation, data is encoded on normalised vectors in
Hilbert spaces. For our purpose, it is enough to say that a Hilbert
space is a vector space over the field of complex numbers. The
smallest space usually considered is the space of {\em qubits}. This
space is the two-dimensional vector space $\mathbb{C}^2$, and comes
with a chosen orthonormal basis denoted by $\{\ket0, \ket1\}$. A
general quantum bit (or qubit) is a normalised vector $\alpha\ket0 +
\beta\ket1$, where $|\alpha|^2 + |\beta|^2=1$. The operations on
qubits that we consider are the {\em quantum gates}, \ie unitary
operations. For our purpose, their interesting property is to be {\em
  linear}.

The language we consider in this paper will be called the {\em
  vectorial lambda-calculus}, denoted by $\lvec$. It is
inspired from {\it Lineal}~\cite{ArrighiDowekRTA08}. This language
admits the regular constructs of lambda-calculus: variables
$x,y,\ldots$, lambda-abstractions $\lambda x.{\ve s}$ and application
$(\ve{s})\,\ve{t}$. It also admits linear combinations of terms:
$\ve{0}$, ${\ve s}+{\ve t}$ and $\alpha\cdot {\ve s}$ are terms. The
scalar $\alpha$ ranges over the ring of complex numbers. As in~\cite{ArrighiDowekRTA08}, it behaves in a
call-by-value oriented manner, in the sense that $(\lambda x.{\ve r})\,({\ve s}+{\ve t})$ first reduces to $(\lambda x.{\ve
  r})\,{\ve s}+(\lambda x.{\ve r})\,{\ve t}$ until {\em basis terms} are reached, at which point beta-reduction applies. The
lambda-binder is not linear with respect to the vectorial
structure: $\lambda x.(\ve{s}+\ve{t})$ is not the same thing as
$\lambda x.{\ve s}+\lambda x.{\ve t}$; in fact abstractions and variables are exactly what is meant by basis terms.

The set of the normal forms of the terms can then be interpreted as a vector space and the
term $(\lambda x.{\ve r})\,{\ve s}$ can be seen as the application of
the linear operator $(\lambda x.{\ve r})$ to the vector~${\ve s}$.
The goal of this paper is to give a formal description of this
intuition at the level of the type system.

\paragraph{Related works and contribution.}
This paper is part of a general research framework aiming at
understanding the relationship between quantum computation and
algebraic lambda-calculi
\cite{AltenkirchGrattageLICS05,ArrighiDowekRTA08,tonder04lambda,ValironQPL10}. The ultimate
goal of this research path is to design a typed language whose terms
can be interpreted both as quantum data and descriptions of quantum
algorithms. The type system would then provide a ``quantum theoretical logic'' and
the language a Curry-Howard isomorphism for quantum computation.

The central question this paper is concerned with is the nature of the
type system to be used. The solution we are proposing is an extension
of two languages designed in \cite{ArrighiDiazcaroQPL09} and
\cite{DiazcaroPetitWoLLIC12}.

The first paper~\cite{ArrighiDiazcaroQPL09} is uniquely concerned with the addition of
scalars in the type system. If $\alpha$ is a scalar and
$\Gamma\vdash\ve t\type T$ is a sequent, $\alpha\cdot\ve t$ is of type
$\alpha\cdot T$. The developed language actually provides a static analysis
tool for {\it probabilistic} computation, when the scalars are taken
to be {\it positive real numbers}. It however fails to address the issue in
this paper: without sums but with negative numbers, the term $\lambda
x.\lambda y.x-\lambda x.\lambda y.y$ is typed with $0\cdot(X\to(X\to
X))$, a type which fails to exhibits the fact that we have a superposition of terms.

The second paper~\cite{DiazcaroPetitWoLLIC12} is concerned with the addition of sums to a
regular type system. In this case, if $\Gamma\vdash
\ve s\type S$ and $\Gamma\vdash\ve t\type T$ are two valid sequents,
$\ve s+\ve t$ is of type $S+T$. However, the language considered is only the {\it
  additive} fragment of {\it Lineal}, it leaves scalars out of the picture.

The paper we present here builds on these two approaches. Its goal is
to characterise the notion of vectors in the vectorial
lambda-calculus. Because of the possible negative or complex
coefficients, this requires to keep track of the `direction' as well
as the `amplitude' of a term.  We propose a type system with both sums
and scalars, reflecting the vectorial structure of the vectorial
lambda-calculus. 
Interestingly enough, combining the two separate features of
\cite{ArrighiDiazcaroQPL09,DiazcaroPetitWoLLIC12} raises subtle novel
issues. In the end we achieve a type system which is such that if $\ve
t$ has type $\sum_i\alpha_i\cdot U_i$, then it must reduce to a $\ve
t'$ of the form $\sum_i\alpha_i\cdot\ve b_i$, where the $\ve b_i$'s
are basis terms.  The resulting language is strongly normalising,
confluent, and features a weak-subject reduction.

\paragraph{Plan of the paper.}
In Section~\ref{sec:language}, we present the language. We discuss the
differences with the original language {\it
  Lineal}~\cite{ArrighiDowekRTA08}. In
Section~\ref{sec:vectorial}, we expose the type system and the problem
arising from the possibility of having linear combinations of types.
Section~\ref{sec:sr} is devoted to subject reduction. We first say
why the usual result is not valid, then we provide a solution and a
candidate subject reduction theorem; the rest of the section is
concerned with the proof of the result.  In Section~\ref{sec:conf}, we prove confluence and strong normalisation for this setting.
Finally we close the paper with some examples in
Section~\ref{sec:examples} and conclusions in
Section~\ref{sec:conclusion}.

\section{The Terms}\label{sec:language}

We consider the untyped language \lvec\ described in
Figure~\ref{fig:Vec}. It is based on {\em Lineal}
\cite{ArrighiDowekRTA08}: terms come in two flavours, basis terms
which are the only ones that will substitute a variable in a
$\beta$-reduction step, and general terms. 

Terms are considered modulo associativity and commutativity of
the operator~$+$, making the reduction into an {\em AC-rewrite system}
\cite{JouannaudKirchnerSIAM86}.
Scalars (notation $\alpha,\beta,\gamma,\dots$) form a ring
$(\Sc,+,\times)$.
The typical ring we consider in the examples is
the ring of complex numbers. In particular, we shall use the shortcut
notation $\ve{s}-\ve{t}$ in place of $\ve{s}+(-1)\cdot\ve{t}$.
The set of free variables of a term is defined
as usual: the only operator binding variables is the
$\lambda$-abstraction.
The operation of substitution on terms (notation~$\ve{t}[{\ve b}/x]$)
is defined in the usual way for the regular lambda-term constructs, by
taking care of variable renaming to avoid capture. For a linear
combination, the substitution is defined as follows: $(\alpha\cdot\ve
t+\beta\cdot\ve r)[{\ve b}/x]=\alpha\cdot\ve t[{\ve b}/x]+\beta\cdot\ve r[{\ve
  b}/x]$.

In addition to $\beta$-reduction, there are fifteen rules stemming from the oriented axioms of vector spaces \cite{ArrighiDowekRTA08}, specifying
the behaviour of sums and products.
A general term $\ve t$ is thought
of as a linear combination of terms $\alpha\cdot\ve r+\beta\cdot\ve
r'$. When we apply $\ve s$ to this superposition, $(\ve s)~\ve t$
reduces to $\alpha\cdot(\ve s)~\ve r + \beta\cdot(\ve s)~\ve r'$.

Note that we need to choose a reduction strategy: we cannot reduce the
term $(\lambda x.(x)~x)~(y+z)$ both to $(\lambda x.(x)~x)~y+(\lambda
x.(x)~x)~z$ and to $(y+z)~(y+z)$. Indeed, the former reduces to
$(y)~y+(z)~z$ whereas the latter reduces to
$(y)~z+(y)~y+(z)~y+(z)~z$. Since this calculus inherits from
\cite{ArrighiDiazcaroQPL09,ArrighiDowekRTA08,DiazcaroPetitWoLLIC12}, we
consider the beta-reduction acting in a call-by-value oriented way (in fact, ``call-by-base'' is a more accurate name).

\begin{figure}[!ht]
\centering
\scalebox{0.8}{\fbox{\begin{tabular}{@{}c@{}}
$$\begin{array}[t]{@{}l@{\hspace{1.5cm}}r@{\ ::=\quad}l@{}}
      \text{\em Terms:} & \ve{t},\ve{r},\ve{u} & \ve{b}~|~(\ve{t})~\ve{r}~|~\ve{0}~|~\alpha\cdot \ve{t}~|~\ve{t}+\ve{r}\\
      \text{\em Basis terms:} & \ve{b} & x~|~\lambda x.\ve{t}
\end{array}$$
\\
\begin{tabular}{p{4cm}p{4cm}p{4cm}}
\emph{Group E:}

  $0\cdot \ve{t}\to\ve{0}$

 $1\cdot \ve{t}\to\ve{t}$

 $\alpha\cdot \ve{0}\to\ve{0}$

 $\alpha\cdot (\beta\cdot \ve{t})\to(\alpha\times\beta)\cdot \ve{t}$

 $\alpha\cdot (\ve{t}+\ve{r})\to\alpha\cdot \ve{t}+\alpha\cdot \ve{r}$

&
\emph{Group F:}

	$\alpha\cdot \ve{t}+\beta\cdot \ve{t}\to(\alpha+\beta)\cdot \ve{t}$

	$\alpha\cdot \ve{t}+\ve{t}\to(\alpha+1)\cdot \ve{t}$

	$\ve{t}+\ve{t}\to(1+1)\cdot \ve{t}$
	
	 $\ve{t}+\ve{0}\to\ve{t}$
\medskip

\emph{Group B:}

 $(\lambda x.\ve{t})~\ve{b}\to\ve{t}[\ve{b}/x]$
&
 \emph{Group A:}

 $(\ve{t}+\ve{r})~\ve{u}\to(\ve{t})~\ve{u}+(\ve{r})~\ve{u}$

 $(\ve{t})~(\ve{r}+\ve{u})\to(\ve{t})~\ve{r}+(\ve{t})~\ve{u}$

 $(\alpha\cdot \ve{t})~\ve{r}\to\alpha\cdot (\ve{t})~\ve{r}$

 $(\ve{t})~(\alpha\cdot \ve{r})\to\alpha\cdot (\ve{t})~\ve{r}$

 $(\ve{0})~\ve{t}\to \ve{0}$

 $(\ve{t})~\ve{0}\to \ve{0}$
\end{tabular}
\\[10ex]
\prooftree\ve{t}\to\ve{s}
\justifies\alpha\cdot \ve{t}\to\alpha\cdot \ve{s}
\endprooftree
\qquad
\prooftree\ve t\to\ve s
\justifies\ve r+\ve t\to \ve r+\ve s
\endprooftree
\qquad
\prooftree\ve t\to \ve s
\justifies(\ve r)~\ve t\to(\ve r)~\ve s
\endprooftree
\qquad
\prooftree\ve t\to \ve s
\justifies(\ve t)~\ve r\to(\ve s)~\ve r
\endprooftree
\qquad
\prooftree\ve t\to \ve s
\justifies\lambda x.\ve t\to\lambda x.\ve s
\endprooftree
\end{tabular}}}
 \caption{Syntax, reduction rules and context rules of \lvec.}
 \label{fig:Vec}
\end{figure}

\paragraph{Relation to other algebraic lambda-calculi.}
Although it is inspired from {\it Lineal}, the language $\lvec$ is closer
 to~\cite{ArrighiDiazcaroQPL09,DiazcaroPerdrixTassonValironHOR10,DiazcaroPetitWoLLIC12}.
Indeed, {\it Lineal} considers some restrictions on the reduction rules, for
example $\alpha\cdot\ve{t}+\beta\cdot\ve{t}\to(\alpha+\beta)\cdot\ve{t}$ is only
allowed when $\ve t$ is a closed normal term.
These restrictions are enforced to ensure confluence in an untyped
calculus. Indeed, consider the following example. Let $\ve Y_\ve
b=(\lambda x.(\ve b+(x)~x))~\lambda x.(\ve b+(x)~x)$. Then $\ve Y_\ve
b$ reduces to $\ve b+\ve Y_\ve b$. So the term $\ve Y_\ve b-\ve Y_\ve
b$ reduces to $\ve 0$, but also reduces to $\ve b+\ve Y_\ve b-\ve
Y_\ve b$ and hence to $\ve b$, breaking confluence. The above
restriction will forbid the first reduction, bringing back confluence.

A series of works
\cite{ArrighiDiazcaroQPL09,DiazcaroPerdrixTassonValironHOR10,DiazcaroPetitWoLLIC12}
has shown that if one considers a typed language enforcing
strong normalisation, one can wave many of the restrictions and
consider a more canonical set of rewrite rules. 
Working with a type system enforcing strong normalisation (as shown in Section~\ref{sec:conf}), we follow this approach.

\paragraph{Booleans in the vectorial lambda-calculus.}
We claimed in the introduction that this language was a candidate language for quantum computation. In this paragraph we show how quantum gates and matrices can be encoded.

First, both in $\lvec$ and in quantum computation one can interpret
the notion of booleans. In the former we can consider the usual
booleans $\lambda x.\lambda y.x$ and $\lambda x.\lambda y.y$ whereas
in the latter we consider the regular quantum bits $\ket{0}$ and
$\ket{1}$.

In $\lvec$, a representation of ${\it if}~{\ve r}~{\it then}~{\ve s}~{\it else}~{\ve t}$ needs to take into
account the special relation between sums and applications. We cannot
directly encode this test as the usual $(({\ve r})\,{\ve s})\,{\ve
  t}$. Indeed, if ${\ve r}$, ${\ve s}$ and ${\ve t}$ were respectively
the terms $\true$, ${\ve s}_1+{\ve s}_2$ and ${\ve t}_1+{\ve t}_2$,
the term $(({\ve r})\,{\ve s})\,{\ve t}$ would reduce to
$((\true)\,{\ve s}_1)\,{\ve t}_1 + ((\true)\,{\ve s}_1)\,{\ve t}_2 +
((\true)\,{\ve s}_2)\,{\ve t}_1 +((\true)\,{\ve s}_2)\,{\ve t}_2$,
then to $2\cdot{\ve s}_1 + 2\cdot{\ve s}_2$ instead of ${\ve s}_1 +
{\ve s}_2$.  We need to ``freeze''
the computations in each branch of the test so that the sum does not
distribute over the application. For that purpose we use the
well-known notion of {\em thunks}: we encode the test as
$\cocanon{(({\ve r})\,\canon{{\ve s}})\,\canon{{\ve t}}}$, where
$\canon{-}$ is the term $\lambda f.-$ with $f$ a fresh, unused term
variable and where $\cocanon{-}$ is the term $(-)\lambda x.x$. The
former ``freezes'' the computation while the latter ``releases''
it. Then the term ${\it if}~\true~{\it then}~({\ve s}_1+{\ve
  s}_2)~{\it else}~({\ve t}_1+{\ve t}_2)$ reduces to the term ${\ve
  s}_1+{\ve s}_2$ as one could expect. Note that this test is linear, in the sense that the term ${\it if}~(\alpha\cdot\true+\beta\cdot\false)~{\it
  then}~{\ve s}~{\it else}~{\ve t}$ reduces to $\alpha\cdot{\ve s} + \beta\cdot{\ve t}$.

This has a striking similarity with the {\it quantum test} that can be
found
\eg in~\cite{AltenkirchGrattageLICS05,ArrighiDowekRTA08,tonder04lambda}.
For example, consider the Hadamard gate {\bf H} sending $\ket0$ to $\frac{\sqrt2}2(\ket0+\ket1)$ and $\ket1$ to
$\frac{\sqrt2}2(\ket0-\ket1)$. If $x$ is a quantum bit, the value
$({\bf H})x$ can be represented as the quantum test $ { {\it
    if}~x~{\it then}~\frac{\sqrt2}2(\ket0+\ket1)~{\it
    else}~\frac{\sqrt2}2(\ket0-\ket1)}$. As developed in
\cite{ArrighiDowekRTA08}, one can simulate this operation in $\lvec$ using the test construction we just
described: $ \cocanon{({\bf H})\,x}$ $=$ $\cocanon{
  ((x)\,\canon{\frac{\sqrt2}2\cdot\true+\frac{\sqrt2}2\cdot\false})\,
  \canon{\frac{\sqrt2}2\cdot\true-\frac{\sqrt2}2\cdot\false} }.  $
Note that the thunks are necessary: the term $((x)\,
(\frac{\sqrt2}2\cdot\true+\frac{\sqrt2}2\cdot\false))\,
(\frac{\sqrt2}2\cdot\true-\frac{\sqrt2}2\cdot\false)$ would reduce to the
term $\frac12(((x)\,\true)\,\true + ((x)\,\true)\,\false +
((x)\,\false)\,\true + ((x)\,\false)\,\false)$, which is fundamentally
different from the term ${\bf H}$ we are trying to emulate.

Of course, with this procedure we can ``encode'' any matrix. If the
space is of some general dimension $n$, instead of the basis elements
$\true$ and $\false$ we can choose the terms $\lambda
x_1.\cdots.\lambda x_n.x_i$'s for $i=1$ to $n$ to encode the basis of
the space.

\section{The Type System}\label{sec:vectorial}

\paragraph{Building the type system.}
Since we are considering a lambda-calculus, we need at least an arrow
type $A\to B$. The terms $\true$ and $\false$ can therefore be typed
in the usual way with $\Bool = X\to(X\to X)$, for a fixed type $X$.
Since the sum ${\frac{\sqrt2}2}\cdot\true +
{\frac{\sqrt2}2}\cdot\false$ is a superposition of terms of type
$\Bool$, one could decide to also type it with the type $\Bool$; in
general, a linear combination of terms of type $A$ would be of type
$A$. But then the terms $\lambda x.(1\cdot x)$ and $\lambda
x.(2\cdot x)$ would both be of the same type $A\to A$, failing to
address the fact that the former respects the norm whereas the latter
does not.

To address this problem, we incorporate the notion of scalars in the
type system: If $A$ is a valid type, the construction $\alpha\cdot A$ is
also a valid type and if the terms $\ve{s}$ and $\ve{t}$ are of type
$A$, the term $\alpha\cdot\ve{s}+\beta\cdot\ve{t}$ is of type
$(\alpha+\beta)\cdot A$. This was achieved in~\cite{ArrighiDiazcaroQPL09} and it allows us to distinguish between
the two functions $\lambda x.(1\cdot x)$ and $\lambda x.(2\cdot x)$:
the former is of type $A\to A$ whereas the latter is of type $A\to
(2\cdot A)$.

Let us now consider the term $\frac{\sqrt2}2\cdot(\true-\false)$. Using
the above addition to the type system, this term should be of type
$0\cdot\Bool$, which is odd in the light of the use we want to
make of it. Indeed, applying the Hadamard gate to this term produces the term
$\false$ of type $\Bool$: the ``amplitude'' of the type (the sum of the squares of the absolute values of the scalars) jumps 
from~$0$~to~$1$.

This time, the problem comes from the fact that the type system does
not keep track of the ``direction'' of a term.  We therefore propose
to go one step further, and to allow sums in types. Provided that
$\True=X\to(Y\to X)$ and $\False=X\to(Y\to Y)$ (with $Y$ another fixed
type), we can type the term $\frac{\sqrt2}2\cdot(\true-\false)$ with
$\frac{\sqrt2}2\cdot(\True-\False)$, which has ``amplitude'' $1$, in the same way that the type of $\false$ has ``amplitude'' $1$.

This type system is also able to type the term ${\bf H}=\lambda
x.((x)\,\canon{\frac{\sqrt2}2\cdot\true+\frac{\sqrt2}2\cdot\false})\,
\canon{\frac{\sqrt2}2\cdot\true-\frac{\sqrt2}2\cdot\false}$, with
$(({\bf I}\to\frac{\sqrt 2}2.(\True+\False)) \to ({\bf
  I}\to\frac{\sqrt 2}2.(\True-\False)) \to T)\to T$
provided that ${\bf I}$ is an identity type of the form $Z\to Z$ and that
$T$ and $Z$ are any fixed types.

Let us try to type the term $\cocanon{({\bf H})\,\true}$. 
This is possible provided
that the fixed type $T$ is equal to ${\bf I}\to\frac{\sqrt
  2}2.(\True+\False)$. If we now want to type the term $\cocanon{({\bf
  H})\,\false}$, the fixed type $T$ needs to be equal to ${\bf
  I}\to\frac{\sqrt 2}2.(\True-\False)$: we cannot type the term $\cocanon{({\bf
  H})\,(\frac{2}{\sqrt2}\cdot\true + \frac2{\sqrt2}\cdot\false)}$ since
there is no possibility to conciliate the two constraints on $T$.

To solve this last problem, we introduce the forall construction in
the type system, making it {\em System~F} alike.  The term ${\bf H}$ can now
be typed with $\forall T.(({\bf I}\to\frac{\sqrt 2}2.(\True+\False))
\to ({\bf I}\to\frac{\sqrt 2}2.(\True-\False)) \to T)\to T$ and the
types $\True$ and $\False$ are updated to be respectively $\forall
XY.X\to(Y\to X)$ and $\forall XY.X\to(Y\to Y)$.  The terms
$\cocanon{({\bf H})\,\true}$ and $\cocanon{({\bf H})\,\false}$ can
both be well-typed with respective types $\frac{\sqrt
  2}2.(\True+\False)$ and $\frac{\sqrt 2}2.(\True-\False)$, as
expected.

\paragraph{The term $\ve0$.}
Let us try to type the term $\ve{0}$. Analogously to what was done for
terms, a natural possibility is to add a special type $\overline{0}$
to type it. This is a reasonable solution that has been used for
example in~\cite{ArrighiDiazcaroQPL09}. In this naive interpretation,
we would have $0\cdot S$ equal to $\overline 0$ and $\overline 0$
would be the unit for the addition on types.

However, consider the following example. Let $\lambda x.x$ be of type $U\to U$
and let $\ve r$ be of type $R$. The term $\lambda x.x + \ve r - \ve r$
is of type $(U\to U) + 0\cdot R$, that is, $(U\to U)$. Now choose $\ve
b$ of type $U$: we are allowed to say that $(\lambda x.x + \ve r - \ve
r)\,\ve b$ is of type $U$. This term reduces to $\ve b + (\ve r)\,\ve
b - (\ve r)\,\ve b$. If the type system is reasonable enough, we
should at least be able to type $(\ve r)\,\ve b$. However, since there is no
constraints on the type $R$, this is difficult to enforce.

The problem comes from the fact that along the typing of $\ve r - \ve
r$, the type of $\ve r$ is lost in the equivalence $\overline 0\equiv
0\cdot R$. The only solution is to distinguish $\overline 0$ from
$0\cdot R$. We can also remove $\overline 0$ altogether, and this is the
choice we make for \lvec: without type $\overline 0$, we do not equate
$T+0\cdot R$ and $T$.

The term $\ve 0$ can be typed with any
type $0\cdot T$, so long as $T$ is inhabited (\ie $\ve 0$ can come
from a reduction of $\ve r - \ve r$ for some term $\ve r$ of type
$T$).

\subsection{Types}

We now give a formal account of the type system. Types are defined in
Figure~\ref{fig:types}. They come in two flavours: {\em unit types}
and general types, that is, linear combinations of types.
Unit types include all types of
\emph{System~F}~\cite[Chapter~11]{GirardLafontTaylor89} and intuitively they are 
used to type basis terms.
The arrow type admits only a unit type in its domain. This is due to
the fact that the 
argument of a
lambda-abstraction can only be substituted by a basis term.
For the same reason, type variables, denoted by~$X, Y$\ldots can
only be substituted by unit types.
The substitution of~$X$ by~$U$ in~$T$ is defined as usual and is
written~$T[U/X]$. For a linear
combination, the substitution is defined as follows: $(\alpha\cdot T+\beta\cdot R)[U/X]=\alpha\cdot T[U/X]+\beta\cdot R[U/X]$.
We also use the vectorial notation~$T[\vec{U}/\vec{X}]$
for~$T[U_1/X_1]\cdots[U_n/X_n]$ if~$\vec{X}=X_1,\dots,X_n$
and~$\vec{U}=U_1,\dots,U_n$, and also $\forall \vec X$ for $\forall X_1\dots X_n=\forall X_1.\dots.\forall X_n$.

We define an equivalence relation~$\equiv$ on types as the least
congruence such that
 $1\cdot T\equiv T$,
$\alpha\cdot T+\beta\cdot T\equiv(\alpha+\beta)\cdot T$,
$\alpha\cdot (\beta\cdot T)\equiv (\alpha\times\beta)\cdot T$,
$T+R\equiv R+T$,
$\alpha\cdot T+\alpha\cdot R\equiv\alpha\cdot (T+R)$,
$T+(R+S)\equiv (T+R)+S$.

This makes the types into a weak module over the scalars: they almost
form a module apart from the fact that there is no neutral element for the
addition. Note that although we do not have any
special type $\overline 0$ (as discussed at the beginning of the section), we
do have $0\cdot T$; however $0\cdot T$ is not the neutral element of the addition
on types.

We may use the summation ($\sum$) notation without ambiguity, due to the associativity and commutativity equivalences of $+$.

\begin{figure}[!ht]
\centering
\scalebox{0.8}{\fbox{\begin{tabular}{@{}c@{}}
$$\begin{array}[t]{l@{\hspace{1.5cm}}r@{\ ::=\quad}l}
      \text{\em Types:} & T,R,S & U~|~\alpha\cdot T~|~T+R\\
      \text{\em Unit types:} & U,V,W & X~|~U\to T~|~\forall X.U
\end{array}$$
\\
\\
$$
\prooftree
\justifies\Gamma, x\type{U}\vdash x\type{U}
\using ax
\endprooftree
\qquad
\prooftree\Gamma\vdash\ve{t}\type T
\justifies\Gamma\vdash\ve{0}\type 0\cdot T
\using 0_I
\endprooftree
\qquad
\prooftree\Gamma, x\type{U} \vdash\ve{t}\type T
\justifies\Gamma \vdash \lambda x.\ve{t}\type{U}\to T
\using\to_I
\endprooftree$$
\\
\\
$$
\prooftree\Gamma \vdash\ve{t}\type\!\!\sui{n}\alpha_i\cdot \forall\vec{X}.(U\to T_i) \qquad \Gamma\vdash\ve{r}\type\!\!\suj{m}\beta_j\cdot {V_j}\qquad
{
\forall V_j,\exists\vec{W}_j, U[\vec{W}_j/\vec{X}]=V_j
}
\justifies\Gamma \vdash(\ve{t})~\ve{r}\type\sui{n}\suj{m} \alpha_i\times\beta_j\cdot {T_i[\vec{W}_j/\vec{X}]}
\using\to_E
\endprooftree$$
\\
\\
$$
\prooftree\Gamma\vdash\ve{t}\type \sui{n}\alpha_i\cdot U_i\quad{X\notin FV(\Gamma)}
\justifies\Gamma\vdash\ve{t}\type\sui{n}\alpha_i\cdot \forall X.U_i
\using \forall_I
\endprooftree
\qquad
\prooftree\Gamma\vdash\ve{t}\type \sui{n}\alpha_i\cdot \forall X.U_i
\justifies\Gamma\vdash\ve{t}\type \sui{n}\alpha_i\cdot U_i[V/X]
\using \forall_E
\endprooftree$$
\\
\\
$$
\prooftree\Gamma\vdash\ve{t}\type T
\justifies\Gamma\vdash\alpha\cdot \ve{t}\type\alpha\cdot T
\using\alpha_I
\endprooftree
\qquad
\prooftree\Gamma\vdash\ve{t}\type T\qquad\Gamma\vdash\ve{r}\type R
\justifies\Gamma\vdash\ve{t}+\ve{r}\type T+R
\using +_I
\endprooftree
\qquad
\prooftree\Gamma\vdash\ve{t}\type T\qquad T\equiv R
\justifies\Gamma\vdash\ve{t}\type R
\using\equiv
\endprooftree
$$
\end{tabular}}}
 \caption{Types and typing rules of \lvec.}
 \label{fig:types}
\end{figure}

The following lemmas give some properties of the equivalence
relation. Types are linear combinations of unit types
(Lemma~\ref{lem:typecharact}). Finally, the equivalence is well-behaved with respect to
type constructs (Lemma~\ref{lem:equivforall}).

\begin{lemma}[Types characterisation]\label{lem:typecharact}
 For any type $T$, there exist $n\in\mathbb{N}$, $\alpha_1,\dots,\alpha_n\in\Sc$ and unit types $U_1,\dots,U_n$ such that $T\equiv\sui{n}\alpha_i\cdot U_i$.
\end{lemma}
\begin{proof}
 Structural induction on $T$. If $T$ is a unit type, take $\alpha=n=1$ and so $T\equiv\sui{1}1\cdot U=1\cdot U$. If $T=\alpha\cdot T'$, then by the induction hypothesis $T'\equiv\sui{n}\alpha_i\cdot U_i$, so $T=\alpha\cdot T'\equiv\alpha\cdot \sui{n}\alpha_i\cdot U_i\equiv\sui{n}(\alpha\times\alpha_i)\cdot U_i$. If $T=R+S$, then by the induction hypothesis $R\equiv\sui{n}\alpha_i\cdot U_i$ and $S\equiv\suj{m}\beta_j\cdot V_j$, so $T=R+S\equiv\sui{n}\alpha_i\cdot U_i+\suj{m}\beta_j\cdot V_j$.
\end{proof}

\begin{lemma}[Equivalence $\forall_I$]\label{lem:equivforall}~
\begin{enumerate}
 \item\label{it:equivforall1} 
$\sui{n}\alpha_i\cdot U_i\equiv\suj{m}\beta_j\cdot V_j\Leftrightarrow\sui{n}\alpha_i\cdot \forall X.U_i\equiv\suj{m}\beta_j\cdot \forall X.V_j$.
 \item\label{it:equivforall2} $\sui{n}\alpha_i\cdot \forall
   X.U_i\equiv\suj{m}\beta_j\cdot V_j\Rightarrow\forall V_j,\exists W_j, V_j\equiv\forall X.W_j$.
 \item\label{it:equivforall3} $T\equiv R\Rightarrow T[U/X]\equiv R[U/X]$.
\end{enumerate}
\end{lemma}
\begin{proof}
 Straightforward case by case analysis over the equivalence rules.
\end{proof}

\subsection{Typing Rules}

The typing rules are described in Figure~\ref{fig:types}. 
Contexts are denoted by $\Gamma$, $\Delta$, etc. and are defined as sets $\{x\type U,\dots\}$, where $x$ is a term variable appearing only once in the set, and $U$ is a unit type.
The axiom
($ax$) and the arrow introduction rule ($\to_I$) are the usual
ones. The rule ($0_I$) to type the term $\ve 0$ takes into account
the discussion at the beginning of Section~\ref{sec:vectorial}. This rule
also ensures that the type of $\ve 0$ is inhabited, discarding
problematic types like $0\cdot \forall X.X$. Any sum of typed terms can be
typed using Rule $(+_I)$. Similarly, any scaled typed term can be typed
with $(\alpha_I)$. Rule $(\equiv)$ ensures that equivalent types can
be used to type the same terms. Finally, the particular form of the
arrow-elimination rule ($\to_E$) is due to the rewrite rules in
group~A that distribute sums and scalars over application. 

The need and use of this complicated arrow elimination can be
illustrated by three examples.

\begin{example}\rm
  Rule $(\to_E)$ is easier to read for trivial linear
  combinations. It states that provided that $\Gamma\vdash \ve s\type
  \forall X.U\to S$ and $\Gamma\vdash \ve t\type V$, if there exists
  some type $W$ such that $V=U[W/X]$, then since the sequent
  $\Gamma\vdash \ve s\type V\to S[W/X]$ is valid, we also have
  $\Gamma\vdash (\ve s)\,\ve t:S[W/X]$.
\end{example}

\begin{example}\rm
  Consider the terms $\ve b_1$ and $\ve b_2$, of respective types
  $U_1$ and $U_2$. The term $\ve b_1 + \ve b_2$ is of type
  $U_1+U_2$. We would reasonably expect the term $(\lambda x.x)\,(\ve
  b_1 + \ve b_2)$ to be also of type $U_1 + U_2$. This is the case
  thanks to Rule $(\to_E)$. Indeed, type the term $\lambda x.x$ with the
  type $\forall X.X\to X$ and we can now apply the rule.
\end{example}

\begin{example}\label{ex:3}\rm
  A slightly more evolved example is the projection of a pair of
  elements. It is possible to encode in {\em System F} the notion of pairs
  and projections: $\langle \ve b, \ve c\rangle = \lambda x.((x)~\ve
  b)~\ve c$, $\langle \ve b', \ve c'\rangle = \lambda x.((x)~\ve
  b')~\ve c'$, $\pi_1 = \lambda x.(x)~(\lambda y.\lambda z.y)$ and
  $\pi_2 = \lambda x.(x)~(\lambda y.\lambda z.z)$. Provided that $\ve
  b$, $\ve b'$, $\ve c$ and $\ve c'$ have respective types $U$, $U'$,
  $V$ and $V'$, the type of $\langle \ve b, \ve c\rangle$ is $\forall
  X.(U\to V\to X)\to X$ and the type of $\langle \ve b',
  \ve c'\rangle$ is $\forall X.(U'\to V'\to X)\to X$. The term $\pi_1$
  and $\pi_2$ can be typed respectively with $\forall XYZ.((X\to Y\to
  X)\to Z)\to Z$ and $\forall XYZ.((X\to Y\to Y)\to Z)\to Z$.
    The term $(\pi_1 + \pi_2)\,(\langle \ve b, \ve c\rangle + \langle
  \ve b', \ve c'\rangle)$ is then typable of type $U+U'+V+V'$, thanks
  to Rule $(\to_E)$. Note that this is consistent with the rewrite
  system, since it reduces to $\ve b + \ve c +
  \ve b' + \ve c'$.
\end{example}

\section{Subject Reduction}\label{sec:sr}

Since the terms of $\lvec$ are not explicitly typed, we are bound to
have sequents such as $\Gamma\vdash\ve{t}\type T_1$ and
$\Gamma\vdash\ve{t}\type T_2$ with distinct types $T_1$ and $T_2$
for the same term $\ve t$.
Using Rules
$(+_I)$ and $(\alpha_I)$ we get the valid typing judgement
$\Gamma\vdash\alpha\cdot \ve t+\beta\cdot \ve t\type\alpha\cdot
T_1+\beta\cdot T_2$. Given that $\alpha\cdot \ve t+\beta\cdot \ve t$
reduces to $(\alpha+\beta)\cdot \ve t$, a regular subject reduction
would ask for the valid sequent $\Gamma\vdash(\alpha+\beta)\cdot \ve
t\type\alpha\cdot T_1+\beta\cdot T_2$.  Since in general we do not
have $\alpha\cdot T_1+\beta\cdot T_2\equiv(\alpha+\beta)\cdot
T_1\equiv(\alpha+\beta)\cdot T_2$, we need to find a way around
this.

A first natural solution could be by using the notion of principal
types. However, since our type system can be seen as an extension of {\em System~F},
the usual examples for the absence of principal types apply to our
settings: we cannot rely on that.

A second potentially natural solution could be to ask for the sequent
$\Gamma\vdash(\alpha+\beta)\cdot \ve t\type\alpha\cdot T_1+\beta\cdot
T_2$ to be valid. If we force this typing rule into the system, it
seems to solve the problem but then the type of a term becomes pretty
much arbitrary: with typing context $\Gamma$, the term
$(\alpha+\beta)\cdot\ve t$ could be typed with any combination
$\gamma\cdot T_1 + \delta\cdot T_2$, when $\alpha+\beta=\gamma+\delta$.

The approach we favour in this paper is by using a notion of order on
types. The order, denoted with $\sqsubseteq$, will be chosen so that
the factorisation rules make the
types of terms smaller according to the order. We will ask in
particular that $(\alpha+\beta)\cdot T_1\sqsubseteq\alpha\cdot
T_1+\beta\cdot T_2$ and $(\alpha+\beta)\cdot T_2\sqsubseteq\alpha\cdot
T_1+\beta\cdot T_2$ whenever $T_1$ and $T_2$ are types for the same
term. 
This approach can also be extended to solve a second pitfall coming the
rule ${\ve t} + \ve0 \to \ve t$. Indeed, although $x:X\vdash x + \ve0
: X+0\cdot T$ is well-typed for any inhabited $T$, the sequent
$x:X\vdash x:X+0\cdot T$ is not valid in general. We therefore extend
the ordering to also allow $X\sqsubseteq X+0\cdot T$.

\subsection{An Ordering Relation on Types.}

We start with another relation $\prec$ inspired from
\cite{Barendregt92}. This relation can be deduced from Rules
$(\forall_I)$ and $(\forall_E)$ as follows: write $T\prec R$ if either
$T\equiv\sui{n}\alpha_i\cdot U_i$ and $R\equiv\sui{n}\alpha_i\cdot
\forall X.U_i$ or $T\equiv\sui{n}\alpha_i\cdot \forall X.U_i$ and
$R\equiv \sui{n}\alpha_i\cdot U_i[V/X]$. We denote the reflexive
(with respect to $\equiv$) and transitive closure of $\prec$ with $\preceq$.
The relation $\preceq$ admits a subsumption lemma.

\begin{lemma}[$\preceq$-subsumption]\label{lem:subsumption}
  For any context $\Gamma$, any term $\ve t$ and any types $T, R$ such that $T\preceq R$ and no free type variable in $T$ occurs in $\Gamma$. Then 
  $\Gamma\vdash\ve{t}\type T$ implies $\Gamma\vdash\ve{t}\type R$.
\end{lemma}
\begin{proof}
 One can assume $\exists S_1,\dots,S_n~/~T\equiv S_1 \prec S_2 \prec
 \cdots \prec S_n \equiv R$ (if not, there must be an equivalence
 instead, so the lemma would hold due to the $\equiv$-rule). So
 for all $i$ one has $S_i\equiv\suj{n}\alpha_j{\cdot} U^i_j$, thus
 $\Gamma\vdash\ve{t}\type \suj{n}\alpha_j{\cdot} U^i_j$ and using
 $(\forall_E)$ or $(\forall_I)$, we get $\Gamma\vdash\ve{t}\type
 \suj{n}\alpha_j{\cdot} U^{i+1}_j$. Since $\suj{n}\alpha_j{\cdot}
 U^{i+1}_j\equiv S_{i+1}$ we finally get $\Gamma\vdash\ve{t}\type
 S_{i+1}$. Repeating the process we eventually reach $\Gamma\vdash\ve{t}\type S_{n}\equiv R$.
\end{proof}

We can now define the ordering relation $\sqsubseteq$ on types
discussed above as the
smallest reflexive transitive relation satisfying the rules:
\begin{enumerate}
 \item $(\alpha+\beta)\cdot T\sqsubseteq\alpha\cdot T+\beta\cdot T'$
   if there are $\Gamma,\ve{t}$ such that $\Gamma\vdash\alpha\cdot \ve{t}\type \alpha\cdot T$ and $\Gamma\vdash\beta\cdot \ve{t}\type \beta\cdot T'$.
\item $T\sqsubseteq T+0.R$ for any type $R$.
\item If $T\preceq R$, then $T\sqsubseteq R$.
\item If $T\sqsubseteq R$ and $U\sqsubseteq V$, then $T+S\sqsubseteq
  R+S$,  $\alpha\cdot T\sqsubseteq\alpha\cdot R$, $U\to T\sqsubseteq
  U\to R$ and $\forall X.U\sqsubseteq\forall X.V$.
\end{enumerate}
Note that the fact that $\Gamma\vdash \ve t\type T$ and $\Gamma\vdash
\ve t\type T'$ does not imply that $\beta\cdot T\sqsubseteq
\beta\cdot T'$. Indeed, although $\beta\cdot T\sqsubseteq 0\cdot
T+\beta\cdot T'$, we do not have $0\cdot T+\beta\cdot
T'\equiv\beta\cdot T'$.
Note also that this ordering is not a subtyping relation. Indeed,
although $\vdash (\alpha+\beta)\cdot \lambda x.\lambda
y.x:(\alpha+\beta)\cdot \forall X.X\to (X\to X)$ is valid and
$(\alpha+\beta)\cdot \forall X.X\to (X\to X)\sqsubseteq \alpha\cdot
\forall X.X\to (X\to X)+\beta\cdot\forall XY.X\to(Y\to Y)$, the
sequent $\vdash (\alpha+\beta)\cdot \lambda x.\lambda y.x:\alpha\cdot \forall
X.X\to (X\to X)+\beta\cdot\forall XY.X\to(Y\to Y)$ is not valid.

\subsection{Weak Subject Reduction}
Let $R$ be any reduction rule from Figure~\ref{fig:Vec}. We denote $\to_R$ a one-step reduction by rule $R$.
A weak version of the subject reduction theorem can be stated as follows.
\begin{theorem}[Weak subject reduction]\label{thm:subjectreduction}
  For any terms $\ve t$, $\ve t'$, any context $\Gamma$ and any type
  $T$, if $\ve{t}\to_R\ve{t}'$ and $\Gamma\vdash \ve t\type T$, then:
  \begin{enumerate}
   \item if $R\notin$ Group F, then $\Gamma\vdash\ve t'\type T$;
   \item if $R\in$ Group F, then
   $\exists S\sqsubseteq T$ such that $\Gamma\vdash\ve t'\type S$ and
   $\Gamma\vdash\ve{t}\type S$.
  \end{enumerate}
\end{theorem}

\noindent How weak is this {\em weak} subject reduction? First, note that the usual subject
reduction result holds for most of the rules. 
Second, Theorem~\ref{thm:subjectreduction} ensures that a term $\ve t$ of a
given type, when reduced, can be typed with a type that is also valid
for the term $\ve t$. 
Third, we can characterise the order relation as follows.
\begin{lemma}[Order characterisation]\label{lem:orderchar} For any type $R$, unit types $V_1,\dots,$ $V_m$ and scalars $\beta_1,\dots,\beta_m$, if $R\sqsubseteq\suj{m}\beta_j\cdot V_j$, then there exist a scalar $\delta$, a natural number $k$, a set $N\subseteq\{1,\dots,m\}$ and a unit type $W\preceq V_k$ such that $R\equiv\delta\cdot W+\sum_{j\in N}\beta_j\cdot V_j$ and $\suj{m}\beta_j=\delta+\sum_{j\in N}\beta_j$.
\end{lemma}
\begin{proof}
 Structural induction on $R$.
\end{proof}

How informative is the type judgement? The following three lemmas express formal relations between the types and their terms.

\begin{lemma}[Scalars, scaling]\label{lem:scalars}\label{cor:scaling}
  For any context $\Gamma$, term $\ve t$, type $T$ and scalar
  $\alpha$, if $\Gamma\vdash\alpha\cdot \ve{t}\type T$, then there
  exists a type $R$ such that $T\equiv\alpha\cdot R$ and if
  $\alpha\neq 0$, $\Gamma\vdash\ve{t}\type R$.  Moreover, if
  $\Gamma\vdash\alpha\cdot \ve{t}\type\alpha\cdot T$, then
  $\Gamma\vdash\ve{t}\type T$.
\end{lemma}
\begin{proof}
 The first part of the Lemma follows by induction on the typing derivation.
 The second part of the Lemma, $\Gamma\vdash\alpha{\cdot}\ve t\type\alpha{\cdot}T\Rightarrow\Gamma\vdash\ve t\type T$, follows as corollary. If $\Gamma\vdash\alpha{\cdot}\ve t\type \alpha{\cdot}T$, we have just proved that there exists $R$ such that $\alpha{\cdot} T\equiv\alpha{\cdot} R$ and $\Gamma\vdash\ve{t}\type R$. It is easy to check that $\alpha{\cdot} T\equiv\alpha{\cdot} R\Rightarrow T\equiv R$, so using rule $\equiv$, $\Gamma\vdash\ve{t}\type T$.
\end{proof}

Lemma~\ref{cor:scaling} is precursor of the generation lemma for
scalars (Lemma~\ref{lem:genLinComb}). However it is more specific since it
assumes a specific type and therefore more accurate in the sense that
it gives a specific type for the inverted rule which is not possible
in the actual generation lemma. 

Lemma~\ref{lem:scalars} excludes the case of scaling by $0$. It is
covered by the following (whose proof is done by induction on the
typing derivation).

\begin{lemma}[Zeros]\label{lem:zeros}
For any context $\Gamma$, term $\ve t$, unit types $U_1,\ldots, U_n$ and scalars $\alpha_1,\ldots,\alpha_n$, if $\Gamma\vdash 0.\ve{t}\type\sui{n}\alpha_i\cdot U_i$, then $\forall i,\,\alpha_i=0$ and there are scalars $\delta_1,\ldots,\delta_n$ such that $\Gamma\vdash\ve{t}\type\sui{n}\delta_i\cdot U_i$.\qed
\end{lemma}

A basis term can always be given a unit type (the proof is also done by induction on the typing derivation).
\begin{lemma}[Basis terms]\label{lem:basevectors}
  For any context $\Gamma$, type $T$ and basis term $\ve{b}$, if
  $\Gamma\vdash\ve{b}\type T$ then there exists a unit type $U$ such
  that $T\equiv U$.\qed
\end{lemma}

In the remainder of this section we provide a few definitions and lemmas that are required in order to prove Theorem~\ref{thm:subjectreduction}.

In the same way that we can change a type in a sequent by an
equivalent one using rule $\equiv$, we can prove that this can also be
done in the context (proof by induction on the typing derivation).

\begin{lemma}[Context equivalence]\label{lem:contextequiv} For any
  term $\ve t$, any context $\Gamma\!=\!(x_i\type\! U_i)_i$ and any type
  $T$, if $\Gamma\vdash\ve{t}\type T$ and $\Gamma'=(x_i\type V_i)_i$
  where $U_i\equiv V_i$, then $\Gamma'\vdash\ve{t}\type T$.\qed
\end{lemma}

The following lemma is standard in proofs of subject reduction for
{\em System~F}-like systems, and can be found, \eg in
\cite[Ch. 4]{Barendregt92}. It
ensures that by substituting type variables for type or term variables
in an adequate manner, the derived type is still valid.

\begin{lemma}[Substitution lemma]\label{lem:substitution} For any term
  ${\ve t}$, basis term $\ve b$, term variable $x$, context $\Gamma$,
  types $T$, $U$, $\vec{W}$ and type variables $\vec{X}$,
  \begin{enumerate}
  \item\label{it:substitution1} if $\Gamma\vdash\ve{t}\type T$, then $\Gamma[U/X]\vdash\ve{t}\type T[U/X]$;
  \item\label{it:substitution2} if $\Gamma,{x}\type U\vdash\ve{t}\type T$, $\Gamma\vdash\ve{b}\type U[\vec{W}\!/\vec{X}]$ and $\vec{X}\notin FV(\Gamma)$, then $\Gamma\vdash\ve{t}[\ve{b}/x]\type T[\vec W\!/\vec X]$.
  \end{enumerate}
\end{lemma}
\begin{proof}
Both results follow by induction on the typing derivation.
\end{proof}

\noindent Proving subject reduction requires the proof that each reduction rule
preserves types. Thus three generation lemmas are required: two
classical ones, for applications (Lemma~\ref{lem:genapp}) and for
abstractions (Lemma \ref{lem:genabs} and
Corollary~\ref{cor:genabs}) and one for linear combinations: sums,
scalars and zero (Lemma~\ref{lem:genLinComb}). The first two lemmas follow by induction on the typing derivation.

\begin{lemma}[Generation lemma (application)]\label{lem:genapp} For
  any terms $\ve t$, $\ve r$, any context $\Gamma$ and any type $T$,
  if $\Gamma\vdash(\ve{t})~\ve{r}\type T$, then there exist natural
  numbers $n,m$, unit types $U, V_1,\dots,V_m$, types $T_1,\dots,T_n$
  and scalars $\alpha_1,\dots,\alpha_n$ and  $\beta_1,\dots,\beta_m$,
  such that $\Gamma\vdash\ve{t}\type\sui{n}\alpha_i\cdot \forall
  \vec{X}.(U\rightarrow T_i)$, $\Gamma\vdash\ve{r}\type\suj{m}
  \beta_j\cdot V_j$, for all $V_j$, there exists $\vec{W}_j$ such that
  $U[\vec{W}_j/\vec{X}]=V_j$ and $\sui{n}\suj{m}
  \alpha_i\times\beta_j\cdot T_i[\vec{W}_j/\vec{X}]\preceq T$.\qed
\end{lemma}

\begin{lemma}[Generation lemma (abstraction)]\label{lem:genabs} For any term variable $x$, term $\ve t$, context $\Gamma$ and type $T$, if $\Gamma\vdash\lambda{x}.\ve{t}\type R$, there exist types $U$ and $T$ such that $U\to T\preceq R$ and $\Gamma,x\type U\vdash\ve{t}\type T$.\qed
\end{lemma}

The following lemma is needed for the proof of Corollary~\ref{cor:genabs}.

\begin{lemma}[Arrows comparison]\label{lem:arrowcomp}
 For any types $T, R$ and any unit types $U,V$, if $V\to R\preceq U\to T$, then there exist $\vec{W},\vec{X}$ such that $U\to T\equiv (V\to R)[\vec{W}/\vec{X}]$.
\end{lemma}
\begin{proof}
 A map $(\cdot)^\circ$ from types to types is defined by
$X^\circ = X$, $(\alpha\cdot  T)^\circ = \alpha\cdot  T^\circ$, $(U\to T)^\circ = U\to T$, $(T+R)^\circ = T^\circ+R^\circ$ and $(\forall X.U)^\circ = U^\circ$.

We need two intermediate results (the first one follows from a structural induction on $T$ and the second one is a case by case analysis on $T\prec R$ using the first result).
\begin{enumerate}
 \item\label{it:ir1} For any type $T$ and unit type $U$, there exists a unit type $V$ such that $(T[U/X])^\circ\equiv T^\circ[V/X]$
 \item\label{it:ir2} For any types $T$, $R$, if $T\preceq R$ then $\exists\vec{U},\vec{X}\ /\ R^\circ\equiv T^\circ[\vec{U}/\vec{X}]$
\end{enumerate}
Proof of the lemma: by definition $U\to T=(U\to T)^\circ$ which by \ref{it:ir2} is equivalent to $(V\to R)^\circ[\vec{W}/\vec{X}]=(V\to R)[\vec{W}/\vec{X}]$.
\end{proof}

\begin{corollary}[of Lemma \ref{lem:genabs}]\label{cor:genabs} For any
  context $\Gamma$, term variable $x$, term $\ve t$, type variables
  $\vec X$ and types $U$ and $T$, if
  $\Gamma\vdash\lambda{x}.\ve{t}\type \forall\vec{X}.(U\to T)$ then the
  typing judgement $\Gamma,x\type U\vdash\ve{t}\type T$ is valid.
\end{corollary}
\begin{proof}
  By Lemma \ref{lem:genabs}, there exist $V,R$ such that $V\to
  R\preceq\forall\vec{X}.(U\to T)$ and $\Gamma,x\type
  V\vdash\ve{t}\type R$. Note that $V\to R\preceq\forall\vec{X}.(U\to
  T)\preceq U\to T$, so by Lemma~\ref{lem:arrowcomp}, there are
  $\vec{W},\vec{Y}$ such that $U\to T\equiv(V\to
  R)[\vec{W}/\vec{Y}]\equiv V[\vec{W}/\vec{Y}]\to R[\vec{W}/\vec{Y}]$
  so $U\equiv V[\vec{W}/\vec{Y}]$ and $T\equiv
  R[\vec{W}/\vec{Y}]$. Also by Lemma~\ref{lem:substitution},
  $\Gamma[\vec{W}/\vec{Y}],x\type V[\vec{W}/\vec{Y}]\vdash\ve{t}\type
  R[\vec{W}/\vec{Y}]$. By Lemma~\ref{lem:contextequiv} and Rule
  $(\equiv)$, $\Gamma[\vec{W}/\vec{Y}],x\type U\vdash\ve{t}\type
  T$. If $\Gamma[\vec{W}/\vec{Y}]\equiv\Gamma$, we are done. Otherwise, $\vec{Y}$ appears free in $\Gamma$. 
  Since $V\to R\preceq U\to T$ and $\Gamma\vdash\lambda x.\ve t\type V\to R$, according to Lemma~\ref{lem:subsumption}, $U\to T$ can be obtained from $V\to R$ as a type for $\lambda x.\ve{t}$: we would need to use Rule $(\forall_I)$; thus $\vec{Y}$ cannot appear free in $\Gamma$, which constitutes a contradiction. So, $\Gamma,x\type U\vdash\ve{t}\type T$.
\end{proof}

\begin{lemma}[Generation lemma (linear combinations)]\label{lem:genLinComb}
    For any context $\Gamma$, scalar $\alpha$, terms $\ve t$ and $\ve r$
  and types $S$ and $T$:
  \begin{enumerate}
   \item if $\Gamma\vdash\ve{t}+\ve{r}\type S$ then there exist types $R$ and
  $R'$ such that $\Gamma\vdash\ve{t}\type R$, $\Gamma\vdash\ve{r}\type
  R'$ and $R+R'\preceq S$;
  \item if $\Gamma\vdash\alpha\cdot \ve{t}\type T$, then there exists a type
  $R$ such that $\alpha\cdot R\preceq T$ and $\Gamma\vdash\alpha\cdot
  \ve{t}\type\alpha\cdot R$;
  \item if $\Gamma\vdash\ve{0}\type T$, then there exists a type $R$ such
  that $T\equiv 0\cdot R$.
    \end{enumerate}
\end{lemma}
\begin{proof}
 All the cases follow by structural induction on the typing derivation.
\end{proof}

\subsection{Proof of Theorem~\ref{thm:subjectreduction}}
We are now ready to prove Theorem \ref{thm:subjectreduction}.

\begin{proof}
Let $\ve t\to_R\ve t'$ and $\Gamma\vdash\ve t\type T$. We proceed by induction. We only give two interesting cases.
\begin{description}
 \item[$R={\alpha\cdot\ve{t}+\beta\cdot\ve{t}}\to{(\alpha+\beta)\cdot\ve{t}}$.]
         Let $\Gamma\vdash\alpha\cdot\ve{t}+\beta\cdot\ve{t}\type T$. Then by
         Lemma~\ref{lem:genLinComb}, there are types $R,S$ such that
         $\Gamma\vdash\alpha\cdot\ve{t}\type R$ and
         $\Gamma\vdash\beta\cdot\ve{t}\type S$ with $R+S\preceq T$.
         By Lemma~\ref{lem:genLinComb}, there exists a type $R'$ such
         that $\alpha. R'\preceq R$
         and $\Gamma\vdash\alpha\cdot\ve t\type\alpha\cdot R'$, and there
         exists a type $S'$ such that $\beta\cdot S'\preceq S$ and $\Gamma\vdash\beta\cdot\ve t\type\beta\cdot S'$.
		\begin{itemize}
		 \item If $\alpha\neq 0$ (or analogously $\beta\neq 0$), then by Lemma~\ref{lem:scalars}, $\Gamma\vdash\ve t\type R'$ and so by $(\alpha_I)$ we conclude $\Gamma\vdash(\alpha+\beta)\cdot\ve t\type(\alpha+\beta)\cdot R'$. Notice that $(\alpha+\beta)\cdot R'\sqsubseteq\alpha\cdot R'+\beta\cdot S'\sqsubseteq R+S\sqsubseteq T$. Also using Rules $(+_I)$ and $(\equiv)$ we conclude $\Gamma\vdash\alpha\cdot \ve t+\beta\cdot \ve t\type (\alpha+\beta)\cdot R'$.
		 
		 \item If $\alpha=\beta=0$, then notice that
                   $\Gamma\vdash 0\cdot \ve t\type 0\cdot R'$ and
                   $0\cdot R'\sqsubseteq 0\cdot R'+0\cdot
                   S'\sqsubseteq R+S\sqsubseteq T$. Using again Rules $(+_I)$ and $(\equiv)$, we conclude $\Gamma\vdash 0\cdot \ve t+0\cdot \ve t\type 0\cdot R'$.
		\end{itemize}
		
\item[$R={(\lambda{x}.\ve{t})~\ve{b}}\to{\ve{t}[\ve{b}/ x]}$.]
      Let $\Gamma\vdash(\lambda{x}.\ve{t})~\ve{b}\type T$. Then by
      Lemma~\ref{lem:genapp}, there exist numbers $n, m$, scalars
      $\alpha_1,\dots,\alpha_n,\beta_1,\dots,\beta_m$, a unit type
      $U$, and general types $T_1,\dots,T_n$ such that
      $\Gamma\vdash\lambda x.\ve{t}\type\sui{n}\alpha_i\cdot
      \forall\vec{X}.(U\to T_i)$ and $\Gamma\vdash\ve{b}\type\suj{m}
      \beta_j\cdot V_j$ with $\sui{n}\suj{m}
      \alpha_i\times\beta_j\cdot T_i[\vec{W}_j/\vec{X}] \preceq T$ and
      where for all $j$, $\vec{W}_j$ is such that $U[\vec{W}_j/\vec{X}]\equiv V_j$.
       
      By Lemma~\ref{lem:basevectors}, $\sui{n}\alpha_i\cdot
      \forall\vec{X}.(U\to T_i)\equiv \forall\vec{X}.(U\to T_i)$ and
      for all $i,k$, $T_i=T_k$. Analogously $\suj{m} \beta_j\cdot
      V_j\equiv V_j$ where for all $j,h$, $V_j=V_h$. So $\sui{n}\alpha_i=1$ and $\suj{m}\beta_i=1$. Then by Rule $(\equiv)$, $\Gamma\vdash\lambda x.\ve{t}\type \forall\vec{X}.(U\to T_i)$,  and $\Gamma\vdash\ve{b}\type V_i$.

      Thus, by Corollary~\ref{cor:genabs}, $\Gamma,x\type
      U\vdash\ve{t}\type T_i$. Notice that $V_i\equiv
      U[\vec{W}_i/\vec{X}]$. By Lemma~\ref{lem:substitution}, we have
      $\Gamma\vdash\ve{t}[\ve{b}/x]\type
      T_i[\vec{W}_i/\vec{X}]$. Since $T_i[\vec{W}_j/\vec{X}]\equiv
      (1\times 1)\cdot T_i[\vec{W}_j/\vec{X}]=
      (\sui{n}\alpha_i)\times(\suj{m}\beta_j)\cdot
      T_i[\vec{W}_j/\vec{X}]=(\sui{n}\suj{m}\alpha_i\times\beta_j)\cdot
      T_i[\vec{W}_j/\vec{X}]$, and since all the $T_i$ are equivalents
      between them, this type is equivalent to
      $\sui{n}\suj{m}\alpha_i\times\beta_j\cdot
      T_i[\vec{W}_j/\vec{X}]\preceq T$. By
      Lemma~\ref{lem:subsumption}, we conclude $\Gamma\vdash\ve{t}[\ve{b}/x]\type T$.\qedhere
\end{description}
\end{proof}

\section{Confluence and Strong Normalisation}\label{sec:conf}
 The language has the usual properties for a typed lambda-calculus: the reduction is locally confluent and the type system enforces strong normalisation. From these two results, we infer the confluence of the rewrite system.
 
\begin{theorem}[Local confluence]\label{thm:localconf}
   For any terms $\ve t$, $\ve r_1$ and $\ve r_2$, if $\ve t\to\ve r_1$ and $\ve t\to\ve r_2$, then there exists a term $\ve u$ such that $\ve r_1\to^*\ve u$ and $\ve r_2\to^*\ve u$. 
\end{theorem}
\begin{proof}
   First, one proves the local confluence of the algebraic fragment of the rewrite system (that is, all the rules minus the beta-reduction). This has been automatised~\cite{BenCoqProof} using COQ~\cite{ManualCOQ}.
   The proof of confluence of the beta-reduction alone is a
   straightforward extension of the proof of confluence of the usual
   untyped $\lambda$-calculus which can be found in many textbooks,
   \eg \cite[Sec. 1.3]{Krivine90}.
   Finally, a straightforward induction entails that the two fragments commute: this entails the local confluence of the whole rewrite system.
\end{proof}

For proving strong normalisation of well-typed terms, we use reducibility candidates, a well-known method described for example in~\cite[Ch. 14]{GirardLafontTaylor89} The technique is adapted to linear combinations of terms. 
 
A {\em neutral term} is a term that is not a lambda-abstraction and
that does reduce to anything. The set of {\em closed neutral terms} is denoted with $\Neutral$. We write $\CT$ for the set of closed terms and $\SN$ for the set of closed, strongly normalising terms. If $\ve t$ is any term, $\Red(\ve t)$ is the set of all terms $\ve t'$ such that $\ve t\to \ve t'$.
It is naturally extended to sets of terms. We say that a set $S$ of closed terms is a reducibility candidate, denoted with $S\in\RC$ if the following conditions are verified:
\begin{description}
 \item[$\RCn_1$] Strong normalisation: $S\subseteq\SN$.
 \item[$\RCn_2$] Stability under reduction: $\ve t\in S$ implies $\Red(\ve t)\subseteq S$.
 \item[$\RCn_3$] Stability under neutral expansion: If $\ve t\in\Neutral$ and $\Red(\ve t)\subseteq S$ then $\ve t\in S$.
 \item[$\RCn_4$] The common inhabitant: $\ve 0\in S$.
\end{description}
\noindent We define the following operations on reducibility
candidates. Let $\bcal{A}$ and $\bcal{B}$ be in $\RC$.
$\bcal{A}\to\bcal{B}$ is the closure of $\{\ve t\in\CT\,|\,\forall\ve
b\in\bcal{A}, (\ve t)\,\ve b\in\bcal{B}\}$ under $\RCn_3$ and
$\RCn_4$, where $\ve b$ is a base term. If $\{\bcal A_i\}_i$ is a
family of reducibility candidates, $\sum_i\bcal{A_i}$ is
the closure of $\{\sum_i\alpha_i\cdot \ve t_i\,|\,\ve t_i\in\bcal{A_i}\}$
under $\RCn_2$ and $\RCn_3$. If there is only one type $A$ in the sum, we
write $1\cdot A$ instead.

\begin{lemma}\label{lem:RCop}
  If $\bcal{A}$, $\bcal{B}$ and all the $\bcal{A}_i$'s are in $\RC$,
  then so are $\bcal{A}\to\bcal{B}$, $\sum_i\bcal{A}_i$ and
  $\cap_i\bcal{A}_i$.
\end{lemma}
\begin{proof}
We need an intermediate result first, showing that
\begin{quote}
if $\{\ve t_i\}_i$ is a family of strongly normalising term, then so is any linear combination of term made of the $\ve t_i$
\end{quote}
Proof of this result: Let $\vec{\ve t}=\ve t_1,\dots,\ve t_n$.  We define the
  {\em algebraic context} $F(\cdot)$ by the following grammar: $F(\vec{\ve t})~::=~\ve t_i\,|\, F(\vec{\ve t}) + F(\vec{\ve t})\,|\,\alpha\cdot F(\vec{\ve t}) \,|\, {\ve 0}.$
  We claim that for all algebraic contexts $F(\cdot)$ and all strongly
  normalising terms ${\ve t}_i$ that are not linear combinations (that
  is, of the form $x$, $\lambda x.\ve r$ or $(\ve s)~\ve r$), the term
  $F(\vec{\ve t})$ is also strongly normalising.
  The claim is proven by induction on $s(\vec{\ve t})$, the sum over
  $i$ of the sum of the lengths of all the possible rewrite sequences
  starting with $\ve t_i$.

\noindent Proof of the lemma:
  \begin{description}
   \item[$\bcal{A}\to\bcal{B}$]
	 $\RCn_1$: Assume that $\ve t\in\bcal{A}\to\bcal{B}$ is not in $\SN$. Then there is an infinite sequence of reductions $({\ve t}_n)_n$ with $\ve t_0 = \ve t$.  So there is an infinite sequence of reduction $((\ve t_n)~\ve b)_n$ starting with $(\ve t)\ve b$, for all base terms $\ve b$.  This contradicts the definition of $\bcal{A}\to\bcal{B}$.
	 $\RCn_2$: We must show that if $\ve t\to\ve t'$ and $\ve
         t\in\bcal{A}\to\bcal{B}$, then $\ve
         t'\in\bcal{A}\to\bcal{B}$. Let $\ve t$ such that for all $\ve b\in\bcal{A}$, $(\ve t)~\ve b\in\bcal{B}$. Then by $\RCn_2$ in $\bcal{B}$, $(\ve t')~\ve b\in\bcal{B}$, and so $\ve t'\in\bcal{A}\to\bcal{B}$.  If $\ve t$ is neutral and $\Red(\ve t)\subseteq\bcal{A}\to\bcal{B}$, then $\ve t'\in\bcal{A}\to\bcal{B}$ since $\ve t'\in\Red(\ve s)$. If $\ve t=\ve 0$, it does not reduce.
	 $\RCn_3$ and $\RCn_4$: Trivially true by definition.
   \item[$\sum_i\bcal{A}_i$]
	 $\RCn_1$: If $\ve t\in \{\sum_i\alpha_i.\ve
           t_i\,|\,\ve t_i\in\bcal{A}_i\}$, the result is trivial by
           condition $\RCn_1$ on the $\bcal{A}_i$ and the previous result about linear combination of strongly normalising terms. If $\ve t$ is neutral and $\Red(\ve t)\subseteq\bcal{A}+\bcal{B}$, then $\ve t$ is strongly normalising since all elements of $\Red(\ve t)$ are strongly normalising.
	 $\RCn_2$ and $\RCn_3$: Trivially true by definition.
	 $\RCn_4$: Since $\sum_i0\cdot\ve
           t_i\in\sum_i\bcal{A}_i$, by $\RCn_2$, $\ve 0$ is also in
           the set.
   \item[$\cap_i\bcal{A}_i$]
	 $\RCn_1$: Trivial since for all $i$, $\bcal{A}_i\subseteq SN$.
	 $\RCn_2$: Let $\ve t\in\cap_i\bcal{A}_i$, then for all $i$, $\ve t\in\bcal{A}_i$ and so by $\RCn_2$ in $\bcal{A}_i$, $\Red(\ve t)\subseteq\bcal{A}_i$. Thus $\Red(\ve t)\subseteq\cap_i\bcal{A}_i$.
	 $\RCn_3$: Let $\ve t\in\Neutral$ and $\Red(\ve t)\subseteq\cap_i\bcal{A}$. Then $\forall_i,\,\Red(\ve t)\subseteq A_i$, and thus, by $\RCn_3$ in $\bcal{A}_i$, $\ve t\in\bcal{A}_i$, which implies $\ve t\in\cap_i\bcal{A}_i$.
	 $\RCn_4$: By $\RCn_4$, $\forall i,\,\ve 0\in\bcal{A}_i$, then $\ve 0\in\cap_i\bcal{A}_i$.\qedhere
   \end{description}
\end{proof}

A \emph{single type valuation} is a partial function from type variables to reducibility candidates, that we define as a sequence of comma-separated mappings, with $\emptyset$ denoting the empty valuation:
$\rho:=\,\emptyset\,|\,\rho,X\mapsto\bcal{A}$.
Type variables are interpreted using pairs of single type valuations, that we simply call {\em valuations}, with common domain:
$\rho = (\rho_+,\rho_-)$ with $|\rho_+|=|\rho_-|$.
Given a valuation $\rho=(\rho_+,\rho_-)$, the {\em complementary valuation} $\bar\rho$ is the pair $(\rho_-,\rho_+)$. We write $(X_+,X_-)\mapsto(A_+,A_-)$ for the valuation $(X_+\mapsto A_+, X_-\mapsto A_-)$. A valuation is called \emph{valid} if for all $X$, $\rho_-(X)\subseteq\rho_+(X)$.

To define the interpretation of a type $T$, we use the following
result.

\begin{corollary}[of Lemma~\ref{lem:typecharact}]\label{cor:typedecomp}
  Any type $T$ has a unique canonical decomposition
  $T\equiv\sui{n}\alpha_i\cdot U_i$ such that for all $j,k$, $U_j\not\equiv
  U_k$.
\end{corollary}
\begin{proof}
   By Lemma~\ref{lem:typecharact}, $T\equiv\sui{n}\alpha_i\cdot
   U_i$. Suppose that there exist
  $j,k$ such that $U_j\equiv U_k$. Then notice that $T\equiv
  (\alpha_j+\alpha_k)\cdot U_j+\sum_{i\neq j,k}\alpha_i\cdot U_i$.
  Repeat the process until there is no more $j,k$
  such that $U_j\not\equiv U_k$.
\end{proof}

The interpretation $\denot{T}_\rho$ of a type $T$ in a
valuation~$\rho=(\rho_+,\rho_-)$ defined for each free type variable
of~$T$ is given by: $\denot{X}_\rho=\rho_+(X)$, $\denot{U\to
  T}_\rho=\denot{U}_{\bar\rho}\to\denot{T}_\rho$, $\denot{\forall
  X.U}_\rho=\cap_{\bcal{A}\subseteq\bcal{B}\in\RC}\denot{U}_{\rho,(X_+,X_-)\mapsto(\bcal{A},\bcal{B})}$,
and if $T\equiv\sum_i\alpha_i\cdot U_i$ is the canonical decomposition
of $T$ then 
$\denot{T}_\rho=\sum_i\denot{U_i}_{\rho}$.
From Lemma~\ref{lem:RCop}, the interpretation of any type is a reducibility candidate.

Reducibility candidates deal with closed terms, whereas proving the adequacy lemma by induction requires the use of open terms with some assumptions on their free variables, that will be guaranteed by a context. Therefore we use \emph{substitutions} $\sigma$ to close terms: $\sigma := \emptyset \;|\; (x \mapsto\ve b;\sigma)$, then $\ve{t}_{\emptyset} = \ve{t}$ and $\ve{t}_{x \mapsto \ve b;\sigma} = \ve{t}[\ve b/x]_{\sigma}$.

  Given a context $\Gamma$, we say that a substitution~$\sigma$
  \emph{satisfies}~$\Gamma$ for the valuation~$\rho$
  (notation:~$\sigma\in\denot{\Gamma}_{\rho}$) when~$(x:U) \in \Gamma$
  implies $x_\sigma\in\denot{U}_{\bar\rho}$ (Note the change in
  polarity).  Let $T\equiv\sui{n}\alpha_i\cdot U_i$, such that for all
  $i,j$, $U_i\not\equiv U_j$, which always exists by
  Corollary~\ref{cor:typedecomp}.
A typing judgement $\Gamma\vdash\ve t\type T$, is said to be \emph{valid} (notation $\Gamma\models\ve t\type T$) if for every valuation~$\rho$, and set of valuations $\{\rho_i\}_n$, where $\rho_i$ acts on $FV(U_i)\setminus FV(\Gamma)$, and for every substitution~$\sigma\in\denot{\Gamma}_\rho$, we have $\ve{t}_{\sigma}\in\sum_{i=1}^n\denot{U_i}_{\rho,\rho_i}$.
\medskip

\begin{lemma}[Adequacy Lemma]\label{lem:SNadeq}
Every derivable typing judgement is valid: For every valid sequent $\Gamma\vdash\ve t:T$, we have $\Gamma\models\ve t:T$.\qed
\end{lemma}
\begin{proof}
 The proof uses a few auxiliary lemmas.
 \begin{enumerate}
  \item\label{lem:polar1}   Given a (valid) valuation $\rho=(\rho_+,\rho_-)$, for all types $T$ we have $\denot{T}_{\bar{\rho}}\subseteq\denot{T}_{\rho}$. 
  {\em Proof:} Structural induction on T
  \item\label{lem:polar2} Let $\rho=(\rho_+,\rho_-)$ and $\rho'=(\rho'_+,\rho'_-)$ be two valuations such that $\forall X$, $\rho'_-(X)\subseteq\rho_-(X)$ and $\rho_+(X)\subseteq\rho'_+(X)$. Then for any type $T$ we have $\denot{T}_{\rho}\subseteq\denot{T}_{\rho'}$ and $\denot{T}_{\bar\rho'}\subseteq\denot{T}_{\bar\rho}$.
  {\em Proof:} Structural induction on $T$.
  \item\label{lem:1cdotrc}
  For all reducibility candidates $\bcal A$, $\bcal A\subseteq 1\cdot
  \bcal A$. Moreover, if $\ve b\in1\cdot\bcal A$ is a base term, then
  $\ve b\in\bcal A$. 
  {\em Proof:} For all $\ve t\in\bcal A$, the term $1\cdot\ve t\in1\cdot\bcal A$. Since
  $1\cdot \ve t\to\ve t$, we conclude using $\RCn_2$.
  Now, consider $\ve b\in 1\cdot\bcal A$. We proceed by structural induction on $1\cdot\bcal A$.
  \item\label{lem:sumrc}
  For all reducibility candidates $\{\bcal A_{i,1}\}_{i=1\cdots n_1}$,
  $\{\bcal A_{i,2}\}_{i=1\cdots n_2}$, if $\ve
  s\in\sum_{i=1}^{n_1}\bcal A_{i,k}$ and $\ve
  t\in\sum_{i=1}^{n_2}\bcal A_{i,2}$, then $\ve s+\ve t\in\sum_{k=1,2, i=1\cdots n_k}\bcal A_{i,k}$. 
  {\em Proof:} By structural induction on $\sum_{i=1}^{n_1}\bcal A_{i,1}$ and
  $\sum_{i=1}^{n_2}\bcal A_{i,2}$.
\item\label{lem:applirc}
  Suppose that $\ve s\in\bcal A\to\bcal B$ and $\ve b\in\bcal A$, then
  $(\ve s)\,\ve b\in\bcal B$. 
  {\em Proof:} Induction on the definition of $\bcal A\to\bcal B$.
 \end{enumerate}
 
 \noindent The proof of the adequacy lemma is done by induction on the size of
the typing derivation of $\Gamma\vdash\ve t\type T$, relying on these
results.
\end{proof}

\begin{theorem}[Strong normalisation]\label{th:SN}
  If $\Gamma\vdash\ve t:T$ is a valid sequent, then $\ve t$ is strongly normalising.
\end{theorem}
\begin{proof}
  If $\Gamma$ is the list $(x_i:U_i)_i$, the sequent $\vdash\lambda x_1\ldots x_n.\ve t:U_1\to(\cdots\to(U_n\to T)\cdots)$ is valid. Using Lemma~\ref{lem:SNadeq}, we deduce that for any valuation $\rho$ and any substitution $\sigma\in\denot{\emptyset}_\rho$, we have $\sigma(\ve t)\in\denot{T}_\rho$. By construction, $\sigma$ does nothing on $\ve t$: $\sigma(\ve t) = \ve t$. Since $\denot{T}_\rho$ is a reducibility candidate, $\lambda x_1\ldots x_n.\ve t$ is strongly normalising. Now suppose that $\ve t$ were not strongly normalising. There would be an infinite rewrite sequence of terms  $(\ve t_i)_i$ starting with $\ve t$. But then $(\lambda\vec{x}.\ve t_i)_i$ would then be an infinite rewrite sequence of terms starting with a strongly normalising term: contradiction. Therefore, $\ve t$ is strongly normalising.
\end{proof}

\begin{corollary}[Confluence]
  If $\Gamma\vdash\ve{s}\type S$ is a valid typing judgement and if
  $\ve{s}\to^*\ve{r}$ and $\ve{s}\to^*\ve{t}$, then there exists $\ve{s}'$
  such that $\ve{r}\to^*\ve{s}'$ and $\ve{t}\to^*\ve{s}'$.
\end{corollary}
\begin{proof}
  A rewrite system that is both locally confluent and strongly
  normalising is confluent~\cite{terese03}.
\end{proof}

\section{Expressing Matrices and Vectors}\label{sec:examples}

In this section we come back to the motivating example introducing the
type system and we show how {\lvec} handles the Hadamard gate, and how
to encode matrices and vectors.

With an empty typing context, the booleans 
$\true=\lambda x.\lambda y.x\,$ and $\,\false=\lambda x.\lambda y.y$
can be respectively typed with the types 
$\True=\forall XY.Y\to (Y\to X)\,$ and $\,\False=\forall XY.X\to (Y\to Y)$.
The superposition has the following type $\vdash\alpha\cdot\true+\beta\cdot\false\type\alpha\cdot\True + \beta\cdot\False$. (Note that it can also be typed with $(\alpha+\beta)\cdot \forall X.X\to X\to X$).

With an empty typing context, the linear map $\ve{U}$ sending $\true$
to $a\cdot\true+b\cdot\false$ and $\false$ to
$c\cdot\true+d\cdot\false$ is written as
$\ve U={\lambda x.((x)\canon{a\cdot\true+b\cdot\false})\canon{c\cdot\true+d\cdot\false}}$.
  The following sequent is valid: 
$\vdash\ve{U}:\forall X.((I\to
(a\cdot\True+b\cdot\False))\to(I\to (c\cdot\True+d\cdot\False))\to
X)\to X$.

This is consistent with the discussion in the introduction:
the Hadamard gate is the case $a=b=c=\frac{\sqrt2}2$ and
$d=-\frac{\sqrt2}2$.  One can check that with an empty typing context,
$\cocanon{({\bf U})~\true}$ is well typed of type
$a\cdot\True+b\cdot\False$, as expected since it reduces to
$a\cdot\true+b\cdot\false$.

The term $\cocanon{({\bf H})~\frac{\sqrt2}2\cdot (\true+\false)}$ is
well-typed of type $\True+0\cdot\False$. Since the term reduces to
$\true$, this is still consistent with the subject reduction: we
indeed have
$\True\sqsubseteq\True+0\cdot\False$.

\section{Conclusion} \label{sec:conclusion}
In this paper we define a strongly normalising, confluent,
typed,
algebraic $\lambda$-calculus satisfying a weak subject reduction. The
language allows making arbitrary linear combinations of
$\lambda$-terms $\alpha\cdot \ve{t}+\beta\cdot \ve{u}$. Its {\em
  vectorial} type system is a fine-grained analysis tool describing the
``vectorial'' properties of typed terms: First, it keeps track of the
`amplitude of a term', \ie if $\ve{t}$ and $\ve{u}$ both have the same
type $U$, then $\alpha\cdot \ve{t}+\beta\cdot \ve{u}$ has type
$(\alpha+\beta)\cdot U$. Then it keeps track of the `direction of a
term', \ie if $\ve{t}$ and $\ve{u}$ have types $U$ and $V$
respectively, then $\alpha\cdot \ve{t}+\beta\cdot \ve{u}$ has type
$\alpha\cdot U+\beta\cdot V$. This type system is expressive enough to be able to
type the encoding of matrices and vectors.

The resulting type system has the property that if $\Gamma\vdash\ve
t\type\sum_i\alpha_i\cdot U_i$ then there exists $\ve t'$ such that
$\ve t\to^*\ve t'$ and $\ve t'=\sum_i\alpha_i{\cdot}\ve b_i$, where
each $\ve b_i$ is a basis term of type $U_i$. Such a $\ve t'$ is
obtained by normalising $\ve t$ under all rules but the factorisation
rules. Within such a $\ve t'$ there may be subterms of the form
$\alpha_1{\cdot}\ve b+\alpha_2{\cdot}\ve b$ of type
$\alpha_1{\cdot}V_1+\alpha_2{\cdot}V_2$, which are redexes for the
factorisation rules. Under our type system, the reduct
$(\alpha_1+\alpha_2){\cdot}\ve b$ can be given both the types
$(\alpha_1+\alpha_2){\cdot}V_1$ and $(\alpha_1+\alpha_2){\cdot}V_2$.

The tool we propose in this paper is a first step towards lifting the ``quantumness'' of algebraic lambda-calculi to the level of a type based analysis. It is also a step
towards a ``quantum theoretical logic'' coming readily with a Curry-Howard
isomorphism. The logic we are sketching merges intuitionistic logic
and vectorial structure. It results into a novel and intriguing tool.

The next step in the study of the quantumness of the linear algebraic
lambda-calculus is the exploration of the notion of orthogonality
between terms, and the validation of this notion by means of a compilation
into quantum circuits. The work in~\cite{ValironQPL10} shows
that it is worthwhile pursuing in this direction.

\paragraph{Acknowledgements}
We would like to thank 
Michele Pagani
and
Barbara Petit
for enlightening discussions.
This work was partially supported by the ANR--JCJC project CausaQ and grants from DIGITEO and Région \^Ile-de-France.

\end{document}